# *JWST* observations of the Ring Nebula (NGC 6720) – II. PAH emission

Nicholas Clark,[1]★ Els Peeters,[1,2,3] Nick L. J. Cox,[4,5] Jan Cami,[1,2,3] Michael J. Barlow,[6] Patrick J. Kavanagh,[7] Griet Van de Steene,[8] Mikako Matsuura,[9] Albert Zijlstra,[10] Isabel Aleman,[11] Jeronimo Bernard-Salas,[4,5] Harriet L. Dinerstein,[12] Kay Justtanont,[13] Kyle F. Kaplan,[12] Arturo Manchado,[14,15,16] Raghvendra Sahai,[17] Peter van Hoof,[8] Kevin Volk[18] and Roger Wesson[6,9]

[1]*Department of Physics and Astronomy, University of Western Ontario, London, Ontario N6G 2V4, Canada*
[2]*Institute for Earth and Space Exploration, University of Western Ontario, London, Ontario N6A 5B7, Canada*
[3]*SETI Institute, Mountain View, CA 94043, USA*
[4]*ACRI-ST, Centre d'Etudes et de Recherche de Grasse (CERGA), 10 Av. Nicolas Copernic, F-06130 Grasse, France*
[5]*INCLASS Common Laboratory, 10 Av. Nicolas Copernic, F-06130 Grasse, France*
[6]*Department of Physics and Astronomy, University College London, Gower Street, London WC1E 6BT, UK*
[7]*Department of Experimental Physics, Maynooth University, Maynooth, Co Kildare, Ireland*
[8]*Royal Observatory of Belgium, Ringlaan 3, B-1180 Brussels, Belgium*
[9]*Cardiff Hub for Astrophysics Research and Technology (CHART), School of Physics and Astronomy, Cardiff University, The Parade, Cardiff CF24 3AA, UK*
[10]*Jodrell Bank Centre for Astrophysics, Department of Physics & Astronomy, The University of Manchester, Oxford Road, Manchester M13 9PL, UK*
[11]*Laboratório Nacional de Astrofísica, Rua dos Estados Unidos, 154, Bairro das Nações, Itajubá, MG, CEP 37504-365, Brazil*
[12]*Department of Astronomy, University of Texas at Austin, Austin, TX 78712, USA*
[13]*Chalmers University of Technology, Onsala Space Observatory, SE-439 92 Onsala, Sweden*
[14]*Instituto de Astrofísica de Canarias, E-38205 La Laguna, Tenerife, Spain*
[15]*Departamento de Astrofísica, Universidad de La Laguna, E-38206 La Laguna, Tenerife, Spain*
[16]*Consejo Superior de Investigaciones Científicas (CSIC), E-28014 Madrid, Spain*
[17]*Jet Propulsion Laboratory, California Institute of Technology, Pasadena, CA 91011, USA*
[18]*Space Telescope Science Institute, 3700 San Martin Drive, Baltimore, MD 21218, USA*

## ABSTRACT

Polycyclic aromatic hydrocarbons (PAHs) and carbonaceous dust have been observed in clumpy circumstellar environments, yet their formation and evolutionary pathways in such environments remain elusive. We aim to characterize the PAH emission in a clumpy planetary nebula to decipher their formation and evolution pathways. We obtained *JWST* Near-Infrared Spectrograph (NIRSpec) and Mid-Infrared Instrument (MIRI) integral field unit spectroscopic observations of two individual knots in the Ring Nebula (NGC 6720), a clumpy planetary nebula, and determine the PAH spectral characteristics. We detect the 3.3 and 11.2 μm PAH emission bands in both knots but do not detect PAH emission in the 6–9 μm range. We supplement our data with *Spitzer* Infrared Spectrograph (IRS) Short-Low 1 (SL1) and SL2 data, containing 11.2, weak 6.2, and weak 7.7 μm PAH emission bands. The *JWST* data confirm the unusual profile of the 11.2 μm band, which is very broad and redshifted with respect to typical 11.2 μm PAH profiles. We estimate the PAH population to be largely neutral. The relative integrated surface brightness of the 3.3 and 11.2 μm bands indicates the presence of small-sized PAHs, consisting of $35 \pm 6$ carbon atoms. We find that the PAH emission is concentrated outside of the clumps, in the inter-clump medium, and confirm the existence of enhanced PAH emission in a narrow 'PAH ring' centred on the central star. This morphology suggests that PAHs formed during the Ring Nebula's asymptotic giant branch phase, in the central star's dust-driven wind.

## 1 INTRODUCTION

The life cycle of carbonaceous matter is critical to many astrophysical and astrochemical processes that are key to galaxy evolution, star and planet formation, and, ultimately, the emergence of life (Tielens 2013; McGuire 2018). A key aspect of this life cycle is understanding the formation and evolution of organic molecules in the extended environments of stars near the end of their evolutionary track, such as planetary nebulae (PNe).

Due to their high temperatures and densities compared to the interstellar medium (ISM), asymptotic giant branch (AGB) stars, post-AGB stars, and PNe are chemical factories. They create a variety of complex molecules, many of which are carbonaceous. Large carbon-based molecules, such as polycyclic aromatic hydrocarbons (PAHs) and fullerenes, are particularly abundant in the extended environments of post-AGB stars and PNe (e.g. Tielens 2008; Cami et al. 2010). Hence, AGB stars, post-AGB stars, and PNe are important laboratories of molecular chemistry. It was once thought that carbon-rich AGB stars were the primary source of PAHs in the ISM (e.g. Allamandola, Tielens & Barker 1989); however, this hypothesis is nowadays under debate (e.g. Sandstrom et al. 2010; Matsuura, Woods & Owen 2013). As such, understanding the processes of formation and evolution of molecules in AGB stars and their evolved counterparts is important to understanding the chemical make-up of the ISM (Latter 1991).

★ E-mail: nclark68@uwo.ca

The Ring Nebula, also referred to as NGC 6720 or M 57, is one of the most emblematic PNe in the night sky. The almost complete lack of clumps seen projected on the central cavity supports the interpretation of the shell as an equatorial or toroidal structure, seen approximately pole-on (Bryce, Balick & Meaburn 1994; O'Dell et al. 2013; Kastner et al. 2025). It is relatively nearby at 790 ± 30 pc (Lindegren et al. 2021) with angular dimensions of 85 arcsec by 65 arcsec for the main ring (see fig. 1 of Wesson et al. 2024). Its central star is a hot white dwarf with an effective temperature $T_{\rm eff} = 135\,000$ K and luminosity $L = 310\,{\rm L}_\odot$ (see Wesson et al. 2024, for these and other stellar and circumstellar properties adopted in this paper). PAH emission has been detected in the Ring Nebula (Cox et al. 2016). PAHs were initially associated with carbon-rich (C/O ≥ 1) envelopes of evolved stars; however, several studies report the detection of PAHs in oxygen-rich circumstellar environments (0.6 < C/O < 1; e.g. Sylvester, Barlow & Skinner 1994; Gutenkunst et al. 2008; Guzman-Ramirez et al. 2014). The Ring Nebula is thought to be oxygen-rich. Using collisionally excited lines, a C/O ratio of 0.6 is found (Liu et al. 2004). It is also possible to determine the C/O ratio from optical recombination lines, and this approach gives a C/O ratio of 0.8 (Liu et al. 2004). We adopt a C/O ratio of less than 1.0 for this work.

Using *Spitzer Space Telescope* observations, Cox et al. (2016) suggest that PAHs also exist on the surface of the relatively dense $H_2$ emitting knots/clumps/globules hereby referred to as 'knots' that are seen towards some PNe (e.g. Hora et al. 2006; Matsuura et al. 2009; Manchado et al. 2015). Unfortunately, *Spitzer* lacked the spatial resolution and spectral sensitivity to unequivocally link the PAH emission to these dense clumps. Cox et al. (2016) found that the conditions on the surface of these knots in the Ring Nebula, with high gas temperature and free C, would be optimal for forming large carbonaceous molecules, including PAHs. However, it was speculated that such an *in situ* formation route would not give rise to the specific population of large, neutral PAHs reported. This work will take advantage of the improved spectral resolution of *JWST* over *Spitzer* to shed some light on these mysteries.

This paper presents new *JWST* Near-Infrared Spectrograph (NIRSpec) and Mid-Infrared Instrument (MIRI) spectroscopic observations of individual knots in the extended environment of the Ring Nebula (NGC 6720). It is the second paper in a series of papers on the *JWST* observations of the Ring Nebula (cf. Wesson et al. 2024; Sahai et al. 2025; van Hoof et al., in preparation). Section 2 describes the *JWST* observations, data reduction, and analysis. We present our results in Section 3 and discuss our findings in Section 4. We summarize our results in Section 5.

## 2 OBSERVATIONS AND DATA REDUCTION

### 2.1 *JWST* observations

This paper is based on spectroscopic observations of the Ring Nebula taken with the *JWST* (Gardner et al. 2023), as part of the Cycle 1 General Observers (GO) programme ID 1558 (PI: M. Barlow) using the NIRSpec instrument (Jakobsen et al. 2022) in integral field unit (IFU) mode (Böker et al. 2022) and the MIRI instrument (Wright et al. 2023) in Medium Resolution Spectrometer (MRS) mode.

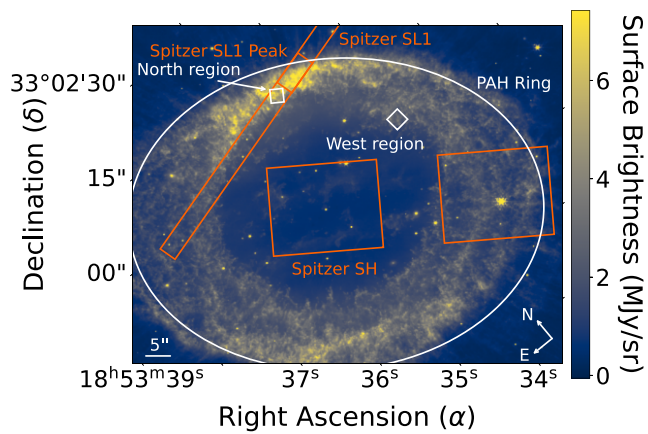

**Figure 1.** NIRCam *F*335*M* image of the Ring Nebula taken from Wesson et al. (2024) with the field of view (FOV) of the *JWST* north and west regions overlaid (white boxes). The 'PAH ring' identified in Wesson et al. (2024) is shown as a white ellipse. The FOV of the *Spitzer*/IRS SH observations used by Cox et al. (2016) and the *Spitzer*/IRS SL1 slit and extraction apertures used in this paper are shown by orange rectangles labelled *Spitzer* SH, *Spitzer* SL1, and *Spitzer* SL1 Peak, respectively. The *Spitzer* SH aperture in the centre of the image corresponds to the central cavity of the Ring Nebula, while the other aperture is located on the outer ring.

*JWST* NIRSpec data are obtained with the $F070LP$-G140M (0.90–1.27 μm), $F100LP$-G140M (0.97–1.89 μm), $F170LP$-G235M (1.66–3.17 μm), and $F290LP$-G395M (2.87–5.27 μm) filter-disperser combinations at medium spectral resolution ($R \sim 1000$) for a total integration time of 583.556 s per segment. The NIRSpec observations cover the spectral range 0.97–5.27 μm nominally across a 30 × 30 grid of 0.1 arcsec 'spaxels'. To quantify the leakage of the Micro-Shutter Array, Leakcal observations are also obtained.

The MIRI/MRS (Wells et al. 2015; Argyriou et al. 2023) comprises four IFUs (channels 1–4), operating across the 4.9–27.9 μm range. These four channels are used in combination with three grating settings (SHORT, MEDIUM, and LONG) to sub-divide the wavelength range into 12 spectral segments, with spectral resolving power ($R = \lambda/d\lambda$) ranging from $\sim$3500 at 5 μm to $\sim$1500 at 27.9 μm (Jones et al. 2023), and the angular resolution from 0.2 to 0.9 arcsec within fields of view of 3.2 arcsec × 3.7 arcsec to 6.6 arcsec × 7.7 arcsec from short to long wavelengths. The total integration time for each sub-channel is 2763.936 s.

The observations are pointed at two knots within the Ring Nebula; i.e. they are centred on RA = $18^{\rm h}53^{\rm m}33^{\rm s}.314$, Dec. = +33°01′49″.90 (referred to as the west region) and RA = $18^{\rm h}53^{\rm m}34^{\rm s}.510$, Dec. = +33°02′09″.11 (referred to as the north region). Both regions are shown in Fig. 1. We also obtained dedicated background observations centred at RA = $18^{\rm h}52^{\rm m}28^{\rm s}.920$ and Dec. = +32°56′29″.76. In this paper, we report on the PAH emission, while van Hoof et al. (in preparation) report on the $H_2$ emission detected in these two regions.

### 2.2 Data reduction

We downloaded the NIRSpec Stage 3 products as processed with the *JWST* pipeline. These NIRSpec data were processed using a recent development version of the *JWST* Calibration Pipeline (Bushouse et al. 2023) with data processing version number 2022_5c, the calibration software version number 1.9.6, and context of calibration reference jwst_1075.pmap. A background and Leakcal correction was applied in the pipeline. Inspection of the extracted spectra showed that the exposures are badly affected by extreme outlier pixel values, which are not flagged in the pipeline processing. These extreme pixel values above 1000 and below $-50$ MJy sr$^{-1}$ per pixel were masked before spectral extraction over the different defined regions.

Inspection of output from an initial pipeline run on our MRS data showed that the exposures are badly affected by cosmic ray showers, which are not flagged in the pipeline processing as they fall below the cosmic ray jump detection threshold. We therefore reduced the MRS data using a recent development version of the *JWST* Calibration Pipeline (Version 1.17.1.dev551 + ga85afaf12) captured on 2025 March 4[1] as it contains a new residual cosmic ray shower correction, which significantly improved our detector images resulting in a much cleaner continuum in extracted spectra. We used versions 12.1.2 and 'jwst_1338.pmap' of the Calibration Reference Data System (CRDS) and CRDS context, respectively.

All of the raw NIRSpec and MIRI files were processed through `Detector1Pipeline` with the default parameters. We accounted for the background using our dedicated 'off' position exposures to create 'master' detector background files for each MRS sub-band, which were subtracted from the corresponding science detector images. The new residual cosmic ray shower correction is part of the `straylight` step in `Spec2Pipeline`.[2] The correction is switched off by default. We switched the correction on and ran `Spec2Pipeline` with all other steps and parameters at their default values to produce flux-calibrated detector images. We constructed spectral cubes for all of the 12 MRS sub-bands using `Spec3Pipeline`, which implements the cube building algorithm described in Law et al. (2023). All spectra extracted from these cubes were processed with a post-pipeline residual fringe correction to account for any residual fringing in the spectra, in particular the high-frequency fringes in channels 3 and 4, thought to originate in the MRS dichroics (Argyriou et al. 2023).

Observing objects with bright emission lines such as PNe produces several detector effects, which can lead to spurious or misleading features in extracted spectra. These effects include spectral contamination due to the 'pull-up/pull-down' electronic cross-talk effect (Dicken et al. 2022), light scattering in the detector across slices and spectral channels, and persistence (Argyriou et al. 2023). We scrutinized both extracted spectra and flux-calibrated detector images to identify spectral features that likely resulted from these effects. We found that all three added spurious spectral features. The scattered light and persistence effects produced spurious, unidentified broad and narrow emission lines, while the pull-up/pull-down affected the continuum in spectral channels with saturated emission lines. Examples of these are shown in Fig. 2. We flagged and ignored the spurious emission lines due to light scattering and persistence in our analysis, which are listed in Table 1. There is currently no correction for the pull-up/pull-down effect. We treated any continuum features in the affected regions with extreme caution.

As we wanted to combine the NIRSpec and MIRI extracted spectra a common region was defined to make sure that the extracted spectra came from the same region. Because the World Coordinate System (WCS) of the NIRSpec and MIRI observations of the north region were misaligned by more than a spaxel, the cubes were first aligned.

---

[1] https://github.com/spacetelescope/jwst
[2] See https://jwst-pipeline.readthedocs.io/en/latest/jwst/straylight/main.html for details.

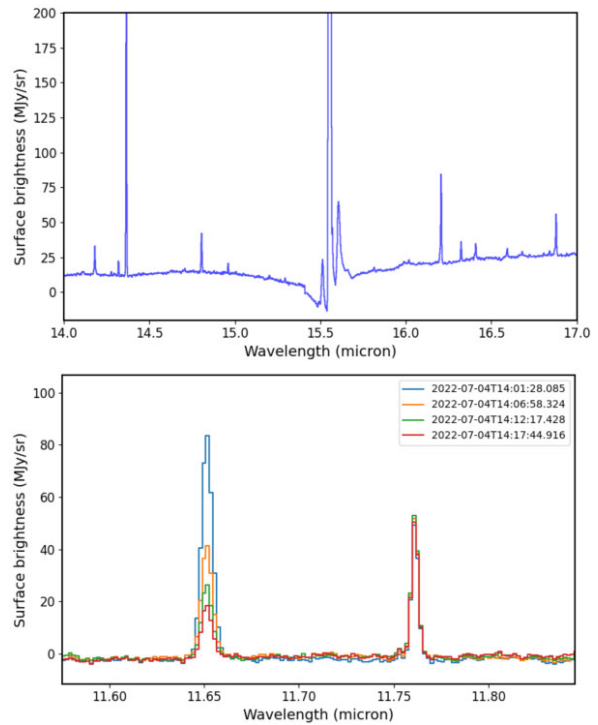

**Figure 2.** Examples of detector effects that affect the continuum and lead to spurious emission features. Top: The 'pull-up/pull-down' leads to the continuum falling below the zero level at wavelengths just before the start of the bright [Ne III] emission line at 15.5 μm. Light scattering between the channel slices produces the spurious, faint emission lines either side of the true [Ne III] line whose peak is above the upper limit of the surface brightness range shown. Bottom: Example of a spurious emission line at ∼11.65 μm. The spectra extracted from a datacube at each of the four dither positions are shown with the timestamps indicated in the legend. The line at ∼11.65 μm fades as the dithers are taken. Note that the real emission line, [Cl IV] at 11.759 00 μm, does not change.

The H$_2$ 5.053 12 μm line is detected by both NIRSpec and MIRI; its flux maps were made from MIRI and NIRSpec data to identify common strong features. The offset between the MIRI and NIRSpec cubes was then corrected using the frame at the peak of the H$_2$ line. We resampled the NIRSpec frame on to the MIRI pixel grid, and then convolved the NIRSpec frame to account for the measured MIRI spatial resolution on the detector. We extract two spectra, using a weighted mean, from two large extraction apertures of size 2.661 arcsec × 2.747 arcsec and 3.015 arcsec × 2.992 arcsec (shown in Fig. 1) encompassing the north and west regions. We use the uncertainty of each pixel for the weights, resulting in extracted spectra for both regions having greatly reduced noise. To remove residual offsets between the different MIRI/MRS sub-channels, we stitched the channels together by adding an offset such that their overlap had the same median. We used channel 1 Short of the MIRI data as the reference spectrum. The resulting spectra are shown in Fig. 3.

### 2.3 Data analysis

To assess the potential presence of PAH emission and, when present, estimate the strength of the PAH bands, we determine the local continuum emission near the 3.3 and 11.2 μm PAH bands. We approximate the continuum as a linear (constant) function for the 3 μm region and for the 11.2 μm PAH feature. We determine the continuum near the 3.3 μm PAH emission based on the observed emission in the 3.1–3.2 μm, and 3.55–3.6 μm range, excluding narrow line emission. To assess the influence of the continuum determination on the 3.3 μm PAH band strength, we also determine two additional continuum levels that are used to determine the upper and lower bounds for the uncertainty on the 3.3 μm PAH feature integrated surface brightness. These extra continuum fits are shown in Fig. A1. For the 11.2 μm PAH feature in the north region, the lower and upper bounds of the continuum are determined based on the median taken from 10.98 to 11.02 μm and from 11.69 to 11.73 μm, respectively. For the west region, the lower and upper bounds of the continuum are determined based on the median from 11.11 to 11.15 μm and from 11.69 to 11.73 μm, respectively. These wavelength ranges are slightly different for the north and west regions to account for the artefact present between 11.0 and 11.13 μm in the west region. This artefact is caused by low-level residual cosmic ray showers affecting part of the detector in some of the background exposures, which produced a slight oversubtraction of the background in the 11–11.13 μm region. The spectra used for comparison have their continuum subtracted using the same method as described above, with the exception of the lower bound for the 11.2 μm feature. Due to the presence of the 11.0 μm PAH feature in these spectra, the lower bound is determined based on the median from 10.80 to 10.84 μm. Emission lines, absorption lines, and artefacts are excluded when determining the continuum. Similar to the 3.3 μm PAH feature continuum, we fit additional continua to serve as the upper and lower bounds for the 11.2 μm PAH feature. The determined continua for the 11.2 μm PAH feature are shown in Figs 3 and A2.

**Table 1.** Spurious lines identified in our MIRI/MRS spectra.

| Field | Wavelength (μm) | MRS band | Cause | Culprit |
| --- | --- | --- | --- | --- |
| North | 4.968 | 1A | Contamination | Residual cosmic ray shower |
| North | 6.365 | 1B | Persistence | $H_2$ 0 − 0 (S7) 5.511 μm |
| West, north | 7.879 | 2A | Persistence | [S IV] 10.510 μm |
| North | 9.268 | 2B | Persistence | $H_2$ 0 − 0 (S4) at 8.025 μm |
| West, north | 10.384 | 3C | Persistence | [Ar III] 8.991 μm |
| West, north | 11.652 | 3A | Persistence | [Ne III] 15.555 μm |
| West, north | 14.805 | 3B | Persistence | [Ne II] 12.823 μm |
| West, north | 17.964 | 3C | Persistence | [Ne III] 15.555 μm |
| West, north | 18.943 | 4A | Persistence | [O IV] 25.890 μm |
| West, north | 21.857 | 4B | Persistence | [S III] 18.713 μm |

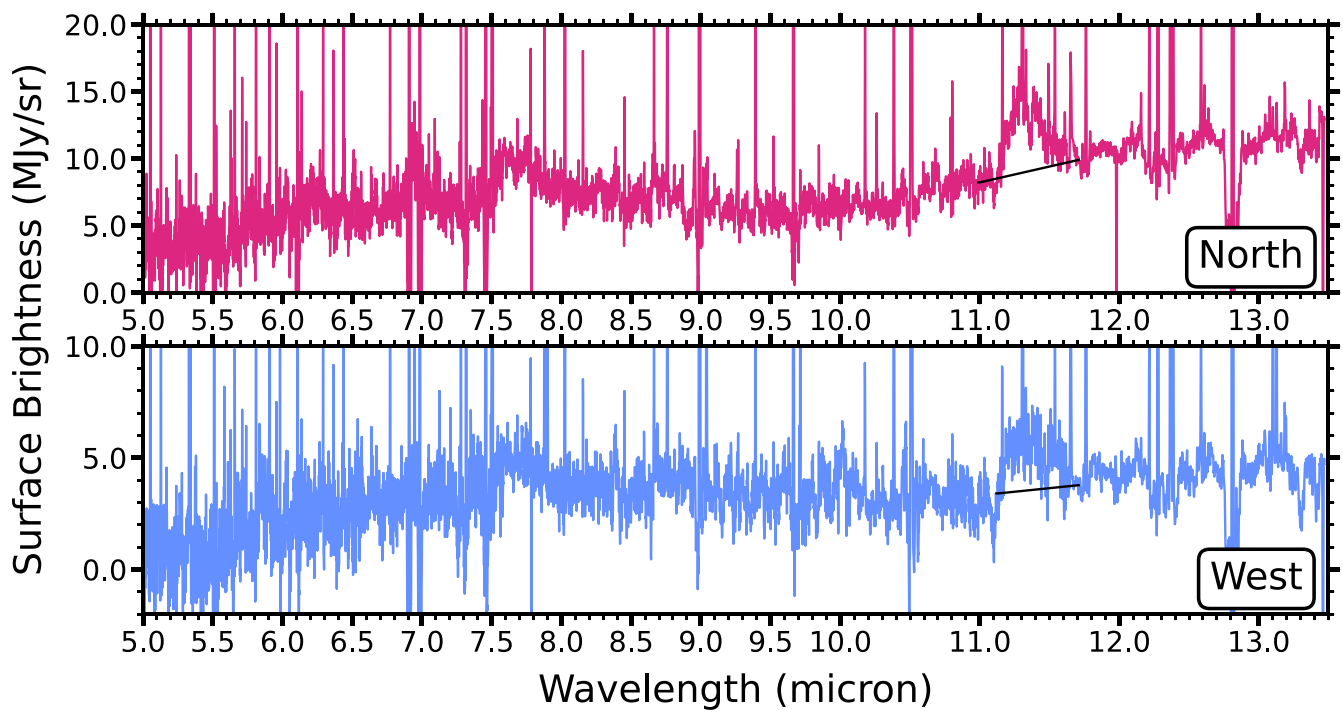

**Figure 3.** The MIRI background-subtracted spectra for the north region (top), and west region (bottom). The continuum fitted under the 11 micron PAH feature is indicated by a black line in both panels (the continuum-subtracted feature is shown in Fig. 5, see analysis for details).

Due to the high density of narrow emission lines located on top of the PAH emission in the 3.2–3.5 μm range, rather than integrating the data itself, Gaussian functions are used to fit the PAH emission in this wavelength range [see Table 2 for their peak position and full width at half-maximum (FWHM)] and we integrated the Gaussians to obtain the 3.3 μm PAH strength (see Fig. 4). Following Peeters et al. (2024), a sum of three Gaussians is used to fully reproduce the PAH profile, with the bulk of the feature fitted by a Gaussian centred on 3.29 μm, with the edges fitted by Gaussian profiles centred on 3.25 and 3.33 μm. In addition to this, a broad Gaussian that represents the PAH plateau feature underneath the 3.3 μm feature is fitted (Peeters et al. 2024). There is potentially excess emission at 3.35–3.5 μm; its detection is hampered by the myriad of blended lines in this wavelength region. We illustrate this potential excess emission by two Gaussians centred near 3.4 μm following Peeters et al. (2024) (see Fig. 4). The abundance of lines on top of the 3.4 μm feature prevents reliable fits with Gaussians; consequently, we did not measure the integrated flux for this feature.

**Table 2.** Peak position, FWHM, and amplitude of the Gaussians used to fit the 3.3 and 3.4 μm features. Note that the peak positions and FWHM of these Gaussians are identical to those used in the Orion Bar (Peeters et al. 2024).

| | North | | |
|---|---|---|---|
| Feature | Peak position[a] | FWHM[a] | Amplitude[b] |
| 3.25 | 3.2465 | 0.0375 | 0.60 |
| 3.29 | 3.2903 | 0.0387 | 2.10 |
| 3.33 | 3.3282 | 0.0264 | 0.20 |
| Plateau | 3.3513 | 0.2438 | 0.35 |
| 3.40 | 3.4031 | 0.0216 | 1.00 |
| 3.42 | 3.4242 | 0.0139 | 0.50 |
| | West | | |
| Feature | Peak position[a] | FWHM[a] | Amplitude[b] |
| 3.25 | 3.2465 | 0.0375 | 0.20 |
| 3.29 | 3.2903 | 0.0387 | 1.10 |
| 3.33 | 3.3282 | 0.0264 | 0.10 |
| Plateau | 3.3513 | 0.2438 | 0.20 |
| 3.40 | 3.4031 | 0.0216 | 0.80 |
| 3.42 | 3.4242 | 0.0139 | 0.40 |

[a]In units of μm.
[b]In units of MJy sr$^{-1}$.

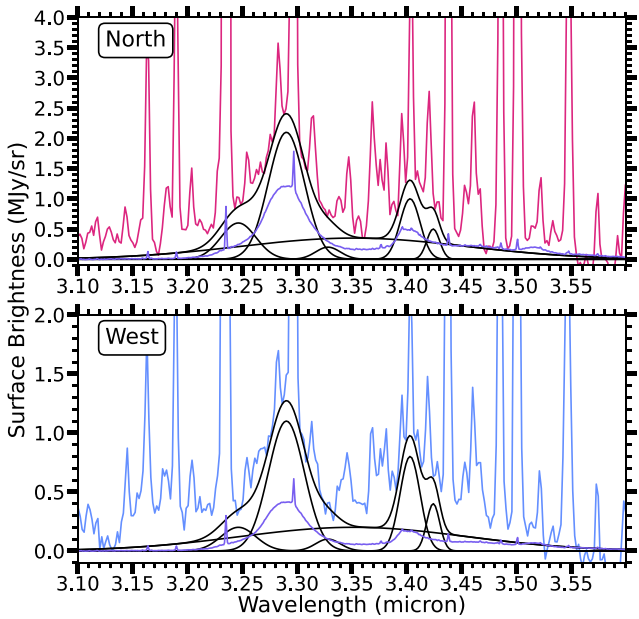

**Figure 4.** Extracted NIRSpec continuum-subtracted spectra. The 3.3 μm PAH feature for the north (top, continuum offset of $-2.2$ MJy sr$^{-1}$), and west (bottom, continuum offset of $-1.0$ MJy sr$^{-1}$) regions is shown. Excess emission with respect to continuum emission and line emission can be seen between 3.35 and 3.45 μm. Three Gaussians are used to fit the 3.3 μm feature in the spectra, and another two are used to fit the 3.4 μm feature and/or excess emission (all Gaussians are shown in black). The continuum-subtracted *JWST* spectrum of the DF3 in the Orion Bar, scaled to match the 11.2 μm PAH feature peak surface brightness, is shown for reference (purple; Peeters et al. 2024). The Orion Bar is chosen because it acts as a benchmark for PAH emission (Chown et al. 2024).

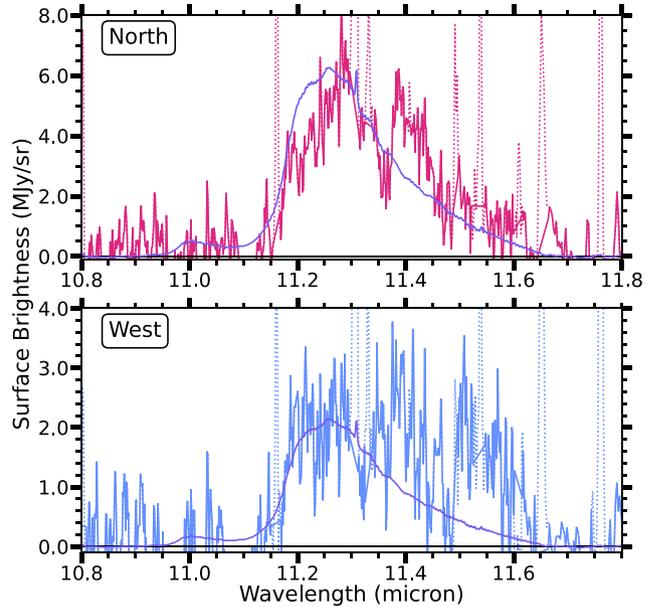

**Figure 5.** The continuum-subtracted 11.2 μm feature for the north (top) and west (bottom) regions. Narrow emission lines that are removed when analysing the 11.2 μm feature are shown by dotted lines. The continuum-subtracted *JWST* spectrum of the DF3 in the Orion Bar, scaled to match the 11.2 μm PAH peak surface brightness (shown in purple), is shown for reference (Chown et al. 2024). The scaling is $1.45 \times 10^{-3}$ and $4.97 \times 10^{-4}$ for the north and west regions, respectively.

The strength of the 11.2 μm PAH band is determined by integrating the continuum-subtracted spectrum between 11.0 and 11.6 μm for the north region. The continuum-subtracted spectrum is integrated between 11.13 and 11.6 μm for the west region, to account for the artefact present between 11.0 and 11.13 μm.

## 3 RESULTS

### 3.1 PAH detections

The MIRI spectra of both the north and west regions are shown in Fig. 3 and the 3.1–3.6 μm and 10.8–12.0 μm continuum-subtracted spectra are shown in Figs 4 and 5, respectively. In both the north and west regions, we clearly detect the 3.3 and 11.2 μm PAH features. The data are too noisy in the 6–9 μm region for the 6.2 or 8.6 μm features to be detected. While there is excess emission between 7.5 and 8 μm, the 7.7 μm PAH complex is typically broader, starting near 7.2 μm. There is also no sign of the 12.0 or 12.7 μm features. We also note excess emission with respect to the continuum and line emission, in the 3.35–3.45 μm range for the two regions.

Cox et al. (2016) reported the detection of PAH emission in the Ring Nebula based on *Spitzer*/IRS Short-High (SH) observations. These authors reported the detection of the 11.2, 12.0, and 12.7 μm PAH features in a 423.2 arcsec$^2$ aperture centred on the Southwest (SW) main dust ring (Fig. 3; fig. 2 of Cox et al. 2016) and no PAH emission in a 423.2 arcsec$^2$ aperture centred on the central hot-gas cavity (see Fig. 1 for the apertures). While we do report the detection of the 11.2 μm PAH, we do not detect the 12.0 and 12.7 μm PAH features. The *Spitzer*/IRS SH aperture and the much smaller *JWST* apertures are centred on different parts of the main dust ring (Fig. 1). Thus, this difference is likely due to the fact that the 12.7 and certainly the 12.0 μm PAH features are much weaker than the 11.2 μm PAH feature. However, we cannot rule out that emission at 12.0 and 12.7 μm is not present in parts of the main dust ring, while emission at 11.2 μm is present throughout the dust ring.

**Table 3.** Integrated surface brightness and peak surface brightness of the *Spitzer*/IRS SL PAH features and their ratios. SL2 was scaled for these values.

| | Surface brightness | |
|---|---|---|
| | Integrated[a] | Peak[b] |
| 6.2 | 3.74 | 2.11 |
| 7.7 | ≤21.15 | ≤7.20 |
| 11.2 | 7.43 | 11.06 |
| 6.2/11.2 | 0.50 | 0.19 |
| 7.7/11.2 | ≤2.85 | ≤0.65 |
| Summary 7.7/11.2 ratios | | |
| Measured | ≤2.85 | |
| From 6.2/11.2 | 0.65 | Section 3.2 |
| Cox et al. (2016) | ≤0.26 | |
| Adopted | $0.65^{2.85}_{0.26}$ | Section 3.2 |

[a]In units of $\times 10^{-8}$ W m$^{-2}$ sr$^{-1}$.
[b]In units of MJy sr$^{-1}$.

We do not conclusively detect PAH emission at 6.2 and 7.7 μm. Thus, the 6.2 and 7.7 μm PAH features are either weak or absent. This is in accordance with the results of Cox et al. (2016) who reported an upper limit of 0.26 for the 7.7/11.2 μm PAH ratio. We investigated *Spitzer*/IRS Short-Low (SL) observations that capture the main dust ring (details on the *Spitzer* observations and data reduction are given in Appendix C). We extracted a spectrum where the emission in the *Spitzer*/IRS SL observation is brightest. This aperture is located just east of the north region FOV discussed here (Fig. 1). Fig. 6 shows that the SL2 order (5.1–7.6 μm) suffers from instrumental effects and has a slope much steeper than that of the SL1 order, not uncommon for weak sources (see e.g. Maragkoudakis et al. 2018). We apply a multiplicative scaling of 0.45 to SL2 to ensure alignment of SL2 with SL1 and SL3 (see Fig. 6). It clearly reveals the presence of the 6.2 μm PAH feature. The mismatch between both orders and the presence of strong emission lines near 7.4–7.5 μm hampers the assessment of the 7.7 and 8.6 μm PAH features. Nevertheless, the 7.7 μm PAH band is likely present and we consider its derived strength to be an upper limit (see Section 3.2 and Table 3).

### 3.2 PAH characteristics

In the north and west regions, the 3.3 μm feature is fitted well by a sum of Gaussians centred at 3.25, 3.29, and 3.33 μm, with the bulk of the emission corresponding to the 3.29 μm Gaussian (Fig. 4). The emission lines present in the data are ignored for the purposes of the fit. Each Gaussian uses the same peak position and FWHM of those used for the fitting of the 3.3 and 3.4 μm features in the atomic zone of the Orion Bar (Peeters et al. 2024; see Table 2). This shows the resemblance of the 3.3 μm feature in these objects.

Excess emission with respect to the continuum and line emission is also detected from 3.35 to 3.45 μm. The wealth of narrow emission lines present on top of this excess emission inhibits the determination of the excess emission's exact profile. However, this excess emission appears to be broader than the typical 3.4 μm band (i.e. when the intensity of the 3.4 μm is significantly smaller compared to the 3.3 μm intensity), in particular for the west region (see Fig. 4). None the less, Gaussians were fitted to the data with the same position and FWHM as was used for fitting the Orion Bar in Peeters et al. (2024) (see Table 2), to demonstrate that there is excess emission not due to blended emission lines present here.

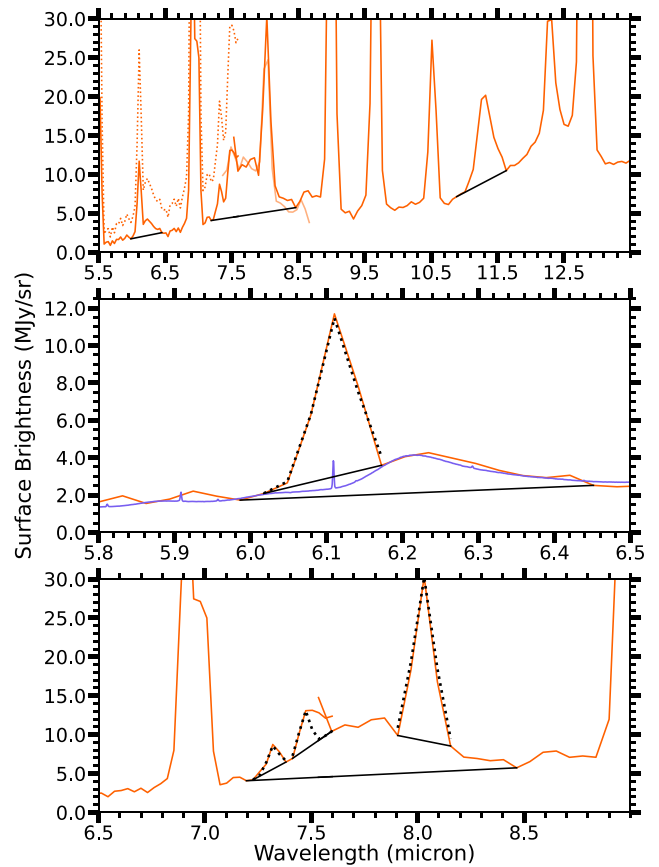

**Figure 6.** *Spitzer*/IRS SL1, SL2, and SL3 spectrum of the Ring Nebula (in orange, light orange, and orange, respectively) at a position just east of the *JWST* FOV of the north region (Fig. 1). The observed (dotted line) and scaled (solid line) SL2 spectrum are shown in the top panel. SL2 is scaled multiplicatively by 0.45 to be aligned with SL1 and SL3. The continua of the 6.2, 7.7, and 11.2 μm PAH features are shown in black. In the middle panel, the 6.2 μm PAH feature is highlighted by scaling the continuum-subtracted spectrum of the Orion Bar (DF3, purple) by 0.0009, then adding the Ring Nebula SL2 continuum for comparison. The H$_2$ 1–0 S(6) line near 6.1 μm in the *Spitzer*/IRS spectrum is fitted with a Gaussian of FWHM set by the spectral resolution (dashed black line) and a local underlying continuum (short solid line). In the bottom panel, the 7.7 μm PAH feature is shown (using the scaled SL2). Emission lines on top of the feature are fitted (dashed black lines on top of local continua and short solid lines).

The 11.2 μm feature is asymmetric and peaks at ∼11.30 μm in the north region and the west region. This is redshifted compared to the 11.2 μm peak position for both 11.2 μm profile classes A and B.[3] In addition, the 11.2 μm feature in both regions is very broad with an FWHM of ∼0.3 μm. Similarly, this is much broader than the width observed for both 11.2 μm profile classes A and B (with typical FWHM of ∼0.17 and ∼0.20 μm, respectively; van Diedenhoven et al. 2004). The dip present at 11.36 μm, ending at 11.40 μm, is likely an artefact and not part of the 11.2 μm feature.

---

[3](PAH features are classified based on their peak position and overall profile, with peak position between 11.20 and 11.24 μm for class A and peak position of ∼11.25 μm for class B; van Diedenhoven et al. 2004)

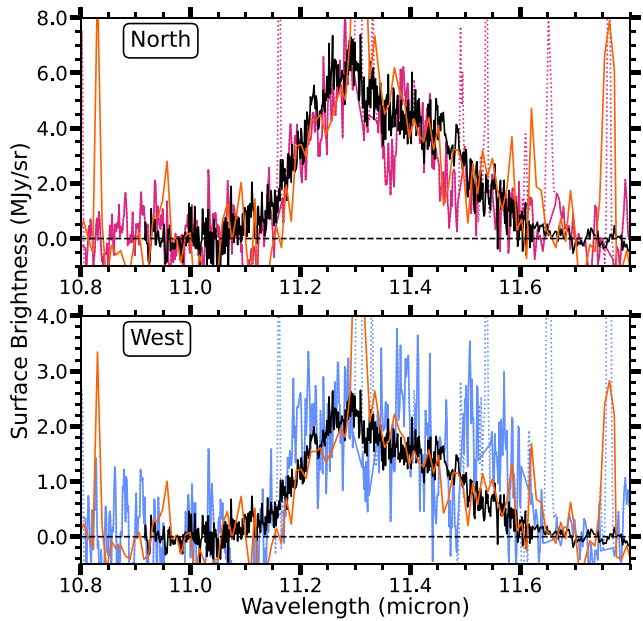
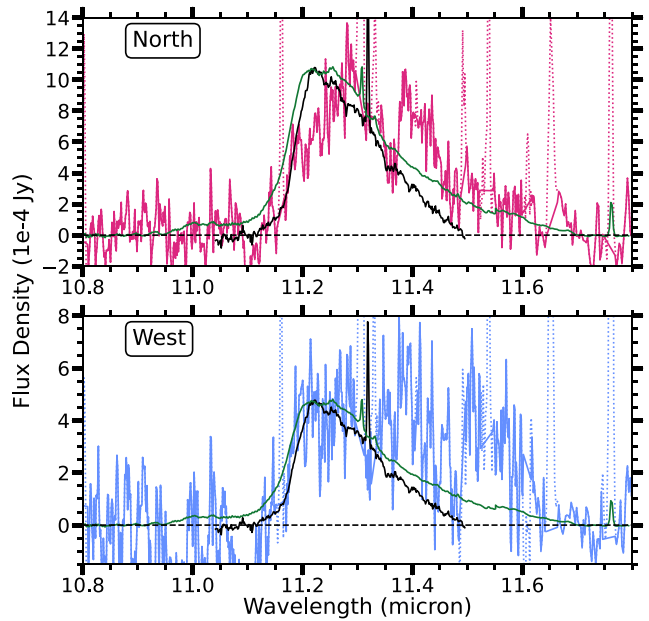

**Figure 7.** Comparison of the continuum-subtracted spectra of the north region (top panel, pink) and west region (bottom panel, blue) in the Ring Nebula (*JWST*) with the continuum-subtracted spectrum of the Horsehead Nebula (*JWST*, shown in black in both panels), and the continuum-subtracted spectrum of the Ring Nebula (*Spitzer*/IRS SH, shown in orange in both panels). These spectra are scaled to match the 11.2 µm peak surface brightness (determined by taking the median from 11.28 to 11.30 µm for the north and west regions). The spectrum of the Horsehead Nebula is scaled by 0.253 and 0.091 for the north and west regions, respectively. The *Spitzer*/IRS SH data are scaled by 1.50 and 0.54 for the north and west regions, respectively.

This is illustrated by its absence in the *Spitzer* data (see Fig. 7). Excluding the dip, the observed 11.2 µm profile in both regions is similar to that observed with *Spitzer*/IRS at a different location in the main dust ring by Cox et al. (2016) (Fig. 7). These authors noted the resemblance with the 11.2 µm profile from the Horsehead Nebula as observed with *Spitzer*/IRS. By extension, this similarity also holds for the *JWST* observations of the north and west regions reported here (Fig. 7), with *JWST* observations of the Horsehead Nebula available in the Barbara A. Mikulski Archive for Space Telescopes (MAST; Programme ID 1192; PI: K. Misselt; for details, see Appendix D).

We also compared our 11.2 µm PAH profiles with those of the PNe SMP LMC 058 (Jones et al. 2023) and NGC 7027 observed with *JWST* (details on these observations and data reduction are given in Appendix D). The 11.2 µm PAH profile is clearly distinct in the Ring Nebula with respect to SMP LMC 058 and NGC 7027 (Fig. 8). The peak of the 11.2 µm feature in the Ring Nebula is redshifted compared to the peak of the 11.2 µm feature of both PNe. As a result, the red wing of 11.2 µm feature in the Ring Nebula is steeper than observed for both comparison PNe, while the blue wing is less steep. Note that both the comparison objects are significantly carbon-rich, which the Ring Nebula is not (Lau et al. 2016; Jones et al. 2023).

We report the integrated and peak surface brightness of the PAH bands for the north and west *JWST* regions in Table 4 and for the *Spitzer*/IRS SL observations in Table 3. We obtain a much higher upper limit for the *Spitzer*/IRS 7.7/11.2 integrated surface brightness ratio than what is reported by Cox et al. (2016, an upper limit of 0.26). This discrepancy may arise from different extraction apertures, data reduction, removal of atomic lines perched on top of the PAH bands, continuum determination, or a combination thereof. However, it is well known that the 6.2 and 7.7 µm PAH strengths correlate very tightly (e.g. Galliano et al. 2008; Bernard-Salas et al. 2009). Hence, based on our observed 6.2/11.2 ratio and a 7.7/6.2 value of $1.3 \pm 0.1$ for PNe (Bernard-Salas et al. 2009), we obtain a 7.7/11.2 ratio of ∼0.65, which is about 2.5 times the upper limit given by Cox et al. (2016) but only 0.2 times the upper limit derived in this paper. Additional data at medium spectral resolution are required to settle this discrepancy. For the remainder of the paper, we will use the 7.7/11.2 ratio obtained from our measured 6.2/11.2 ratio while considering both upper limits as its bounds [i.e. effectively treating the upper limit of Cox et al. (2016) as a lower limit].

**Figure 8.** Comparison of the continuum-subtracted spectra of the north region (top panel, pink) and west region (bottom panel, blue) in the Ring Nebula (*JWST*) with the continuum-subtracted spectrum of the PN SMP LMC 058 (black), and the PN NGC 7027 (green). The flux density of these objects is shown as opposed to the surface brightness, because SMP LMC 058 is an unresolved source. These spectra are scaled to match the 11.2 µm peak flux density. The SMP LMC 058 spectra are scaled by 350 and 126 for the north and west regions, respectively. NGC 7027 is scaled by $1.01 \times 10^{-4}$ and $3.61 \times 10^{-5}$ for the north and west regions, respectively.

**Table 4.** Integrated surface brightness and peak surface brightness of the detected PAH features and their ratios.

|  | North | West | North | West |
| --- | --- | --- | --- | --- |
|  | Surface brightness | | | |
|  | Integrated[a] | | Peak[b] | |
| 3.3 | 3.23 ± 0.30 | 1.56 ± 0.18 | 2.40 ± 0.19 | 1.25 ± 0.23 |
| 11.2 | 3.34 ± 0.83 | 1.71 ± 0.33 | 6.20 ± 0.50 | 2.23 ± 0.50 |
| 3.3/11.2 | 0.97 ± 0.26 | 0.91 ± 0.20 | 0.39 ± 0.04 | 0.56 ± 0.16 |

[a]In units of $\times 10^{-8}$ W m$^{-2}$ sr$^{-1}$.
[b]In units of MJy sr$^{-1}$.

Following Gordon et al. (2022), we estimate the fractional contribution of the 3.3 µm PAH emission to the NIRCam *F*335*M* filter (using the sum of Gaussians reported in Table 2).[4] We obtain a fractional contribution of the PAH emission of 5.4 and 5.1 per cent (corresponding to an absolute surface brightness of 0.31 and 0.13 MJy sr$^{-1}$) for the north and west regions, respectively. This is consistent with Wesson et al. (2024) who reported an upper limit of 4 per cent for the PAH contribution to the *F*335*M* filter for the north region.

---

[4] We find excellent agreement between the absolute flux calibration of NIRCam *F*335*M* and NIRSpec: we obtain a synthetic surface brightness of

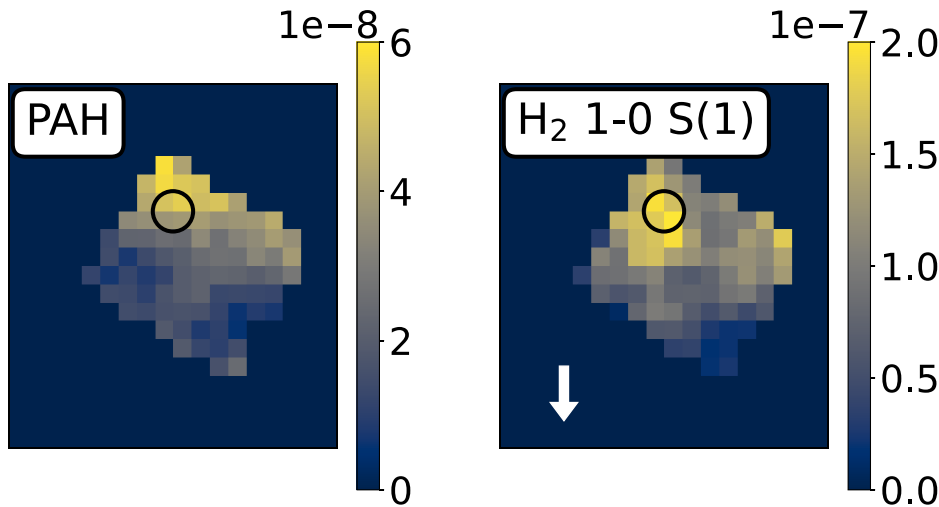

**Figure 9.** The morphology of the 3.3 μm PAH emission (left), and the H$_2$ 1–0 S(1) emission (right) for the north region. The black circle indicates the location of the peak H$_2$ 1–0 S(1) emission. The arrow in the right panel indicates the approximate direction of the central source. The data are regridded to 2 pixels by 2 pixels, in order to reduce noise. The integrated surface brightness displayed on the colourbar is in units of W m$^{-2}$ sr$^{-1}$. The maps on a pixel scale of 0.1 arcsec are shown in Fig. E1.

### 3.3 PAH spatial distribution

We investigate the 3.3 μm PAH feature morphology in the north region of the Ring Nebula.[5] We fit a Gaussian centred on 3.29 μm with FWHM of 0.0387 μm to the 3.3 μm PAH emission assuming a flat, constant continuum. The continuum is estimated by a line of best fit through 3.1–3.2 and 3.45–3.6 μm. Fig. 9 shows the spatial distribution of the 3.3 μm PAH emission and the H$_2$ 1–0 S(1) emission. Appendix E shows the corresponding maps at the original pixel scale of 0.1 arcsec as well as the signal-to-noise ratio maps of these emission features. The 3.3 μm PAH emission peaks furthest away from the central star in the north region. Comparison with the H$_2$ emission reveals both tracers to be weaker in the lower half of the FOV, closest to the star. However, the peak emission of these tracers is quite distinct. In particular, regions where the H$_2$ emission peaks (locally) exhibit decreased PAH emission. In addition, the 3.3 μm PAH peak emission is located at a further distance from the star than the H$_2$ peak emission. Wesson et al. (2024) reported enhanced H$_2$ emission in the dense globules of the Ring Nebula. We refer the reader to van Hoof et al. (in preparation) for a detailed description of the origin of the H$_2$ emission. The spatial displacement of the PAH and H$_2$ emission therefore indicates that the main PAH emission does not originate from the surface of the dense knots.

---

5.46 and 2.45 MJy sr$^{-1}$ for the north and west regions, respectively, using the NIRSpec observations, which agrees very well with the observed NIRCam *F*335*M* surface brightness of 5.55 and 2.63 MJy sr$^{-1}$, respectively.
[5] Unfortunately, the quality of the MIRI/MRS data inhibits studying the morphology of the 11.2 μm feature, and the weaker PAH emission in the west region inhibits a similar analysis of its 3.3 μm feature.

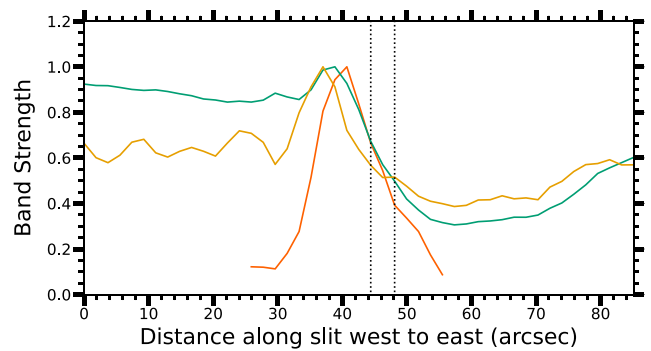

**Figure 10.** The integrated surface brightness of the *Spitzer*/IRS SL1 11.2 μm PAH emission (orange, in units of W m$^{-2}$ sr$^{-1}$), the NIRCam *F*335*M*/*F*300*M* ratio (yellow, in units of MJy sr$^{-1}$), and the MIRI *F*1130*W*/*F*1000*W* ratio (green, in units of MJy sr$^{-1}$) along the *Spitzer*/IRS SL1 slit shown in Fig. 1. All measurements shown are ≥3σ detections. Each curve is scaled to have a peak band strength of 1. Scaling factors are 7.22 × 10$^{-8}$, 2.38, and 1.54, respectively. The vertical dashed lines indicate the location of the *JWST* north region, as it is contained within the *Spitzer*/IRS SL1 slit. The aperture '*Spitzer*/IRS SL1 Peak' seen in Fig. 1 corresponds to 35–45 arcsec along the slit.

In addition, we investigate the PAH morphology based on the 11.2 μm PAH (*Spitzer*/IRS SL1 observations; Fig. 1). For each pixel along the slit, we integrate the continuum-subtracted spectrum from 11.073 to 11.632 μm assuming a linear continuum with anchor points set by the average surface brightness in the 10.825–10.949 and 11.632–11.880 μm range. The 11.2 μm PAH emission along the slit is strongly centred on a very small region of the slit, which is not included in the FOV of the *JWST* north region (Fig. 10). Moreover, Figs 10 and 11 indicate that the peak 11.2 μm PAH emission coincides with the enhanced emission observed in the NIRCam *F*335*M*/*F*300*M* and MIRI *F*1130*W*/*F*1000*W* ratio maps reported by Wesson et al. (2024).

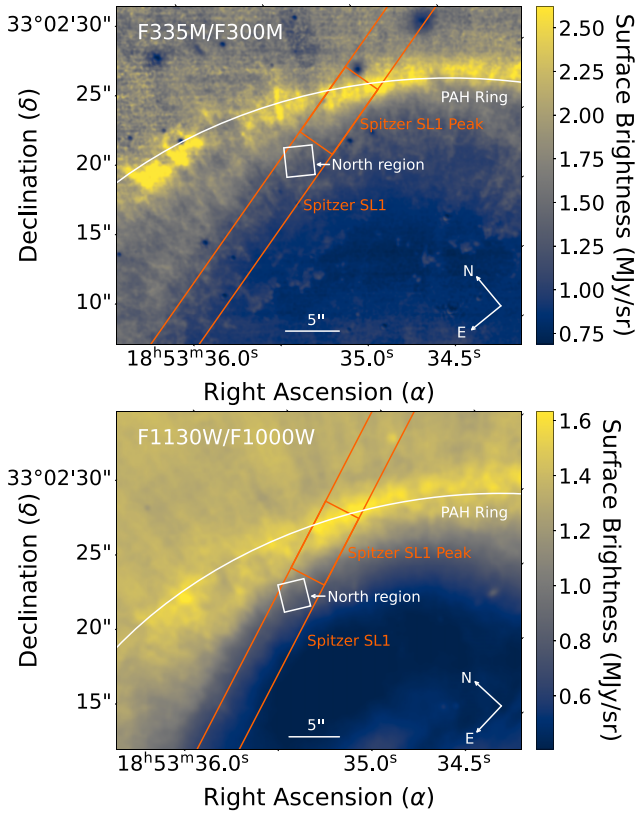

**Figure 11.** The NIRCam *F335M*/*F300M* ratio (top), and the MIRI *F1130W*/*F1000W* ratio (bottom) maps taken from Wesson et al. (2024). The PAH ring is indicated in a white curved line. The *Spitzer*/IRS SL1 slit used in Fig. 10 is indicated by orange lines. The FOV of the *Spitzer*/IRS SL aperture is shown by an orange rectangle, and the *JWST* north aperture is shown by a white rectangle.

## 4 DISCUSSION

### 4.1 Characteristics of the PAH population

The 6.2/11.2 and 7.7/11.2 ratios are commonly used as a tracer for PAH charge (e.g. Allamandola, Hudgins & Sandford 1999; Galliano et al. 2008). The PAH charge balance is set by the ratio of the photoionization rate and the recombination rate or the PAH ionization parameter $\gamma = G_0 T^{0.5}/n_e$ (Bakes & Tielens 1994). Here, $G_0$ is the radiation field strength in Habing units, equal to $1.6 \times 10^{-6}$ W m$^{-2}$, with $T$ the gas temperature, and $n_e$ the electron density. PAH anions and PAH neutrals dominate the PAH charge populations in environments where the PAH ionization parameter $\gamma$ is less than a few $10^2$ and between $\sim 10^3$ and $10^4$, respectively (Bakes, Tielens & Bauschlicher 2001; Sidhu et al. 2022). We emphasize that PAH anions and PAH cations give rise to similar emission characteristics and thus similar 6.2/11.2 and 7.7/11.2 ratios compared to that of PAH neutrals (see e.g. figs 8 and 11 of Sidhu et al. 2022).

Following Croiset et al. (2016) and Schefter et al. (in preparation), we estimate the averaged absorbed photon energy based on blackbody radiation at a temperature equal to the effective temperature of the star ($T_{eff} = 135\,000$ K) for circumcoronene ($C_{54}H_{18}$). We adopt the ultraviolet (UV) absorption cross-section of neutral circumcoronene from Malloci, Mulas & Joblin (2004)[6] and obtain an average absorbed photon energy of 10 eV. No internal extinction is considered here. The diagnostic charge-size diagram for a 10 eV average absorbed photon energy from Maragkoudakis, Peeters & Ricca (2020, their fig. 10) points towards an almost fully neutral PAH population (0–15 per cent cations for a PAH size of about 30–120 carbon atoms) for a 7.7/11.2 ratio of 0.65, whereas its bounds correspond to a fully neutral PAH population (lower bound) and a PAH population consisting of less than 75 per cent cations (upper bound).[7]

[6] http://astrochemistry.oa-cagliari.inaf.it/database/pahs.html

The observed 6.2/11.2 ratio is consistent with environments where the PAH ionization parameter $\gamma$ is between $\sim 10^3$ and $10^4$ (Sidhu et al. 2022), i.e. where PAHs are neutral (Bakes et al. 2001; Sidhu et al. 2022). Combined with the PAH charge fractions derived for the three 7.7/11.2 ratios considered above, the observations preferentially point towards a PAH population that is largely neutral. Given that the Ring Nebula has a value of $G_0$ ranging from 200 to 400 within the ring (Cox et al. 2016) and assuming a gas temperature in the Photodissociation region (PDR; where PAHs reside) of a few 100s K and an electron density fraction of $1.6 \times 10^{-4}$, this implies gas densities of $10^3$ to a few $10^4$ cm$^{-3}$. Such gas densities are typical for PDRs but much lower than those for the dense clumps found in PNe, which is around $10^6$ cm$^{-3}$ (Matsuura et al. 2007). Gas densities of the order of $10^6$ cm$^{-3}$ would imply a PAH ionization parameter $\gamma$ of roughly a few tens, i.e. where the PAH population is dominated by anions (Bakes et al. 2001; Sidhu et al. 2022). This in turn implies a 6.2/11.2 ratio of 1–10 depending on the PAH size and the excitation conditions of the region ($G_0$ and $T_{eff}$; Sidhu et al. 2022), considerably higher than the observed 6.2/11.2 ratio.

The 3.3/11.2 PAH ratio is an excellent tracer of the average size of PAH population (Schutte, Tielens & Allamandola 1993; Ricca et al. 2012; Croiset et al. 2016; Maragkoudakis et al. 2020; Knight et al. 2021). Indeed, smaller PAHs have fewer vibrational modes than larger PAHs ($3N-6$ with $N$ the number of atoms). Consequently, upon absorption of a given UV photon, smaller PAHs will obtain a higher internal temperature as more energy is distributed per mode. As a result, smaller PAHs will thus preferentially emit in the 3.3 $\mu$m band compared to the 11.2 $\mu$m band. Moreover, both bands originate from neutral PAHs eliminating the effect of PAH charge, and bracket a large wavelength range making their ratio most sensitive to changes in the PAH size.

Based on the observed 3.3/11.2 band ratio (Table 4), a purely neutral PAH population, and the diagnostic charge-size diagram for a 10 eV average absorbed photon energy from Maragkoudakis et al. (2020, their fig. 10), we obtain a PAH size of approximately $43 \pm 7$ carbon atoms for both the north and west regions. This diagnostic diagram is based on intrinsic spectra from the NASA Ames PAH infrared (IR) spectral data base computed in the harmonic approximation using a limited basis set (Bauschlicher et al. 2010, 2018; Boersma 2014). Recently, Mackie et al. (2022) and Lemmens et al. (2023) reported that the 3.3 $\mu$m band intensity in these spectra is overestimated relative to the out-of-plane CH bending modes (10–15 $\mu$m) by 34 per cent. Taking this correction factor into account for the diagnostic diagram, we obtain a PAH size of approximately $35 \pm 6$ carbon atoms for both the north and west regions.[8] Implicit in this derivation is the assumption that (1) despite the unusual characteristics of the 11.2 µm PAH band, its (relative) intensity is typical for PAHs, and (2) multiphoton events can be ignored.

---

[7] Anions are not considered in these charge-size diagrams.
[8] Considering a PAH population with 75 percent cations and taking into account this correction factor, we obtain a PAH size of approximately $26 \pm 5$ carbon atoms in both the north and west regions.

Comparison with the Orion Bar templates (Chown et al. 2024; Peeters et al. 2024) clearly indicates that the 3.3 µm band relative to the 11.2 µm band is significantly stronger than in the atomic PDR and the third dissociation front (DF3) of the Orion Bar (Fig. 4; Habart et al. 2024; Peeters et al. 2024). This holds both when considering integrated or peak surface brightness. However, the Orion Bar is subject to a different radiation field and the average absorbed photon energy is lower than in the Ring Nebula (Schefter et al., in preparation). Taking into account the radiation field in the Orion Bar PDR, the PAHs in the Ring Nebula are similar to the smallest PAHs found in the Orion Bar (Lemmens et al. 2023).

**4.2 Location of PAH emission**

The morphology of the *JWST* NIRSpec 3.3 µm PAH emission indicates that the PAH emission does not originate from a dense clump, as traced by $H_2$ emission (Section 3.3). This is further supported by the derived gas density from the observed *Spitzer*/IRS SL 6.2/11.2 ratio, which is typical for PDRs (Section 4.1). While PAH emission from clumps will arise from the clump's surface layers, where gas densities are lower, the 3.3 µm PAH morphology indicates that the PAH emission is too far from the clumps to be on the clump's surface. This clearly points towards an origin in the inter-clump region. In addition, we confirm the existence of a ring of enhanced PAH emission reported by Wesson et al. (2024). This PAH ring encompasses the region with enhanced $H_2$ emission. Furthermore, as the molecular ring was probably not ionized and the clumps may be recombining (Wesson et al. 2024), it may suggest that recombination of the ionized gas does not lead to PAH formation. Hence, the PAH morphology likely points towards the formation of PAHs during the Ring Nebula's AGB phase, in the central star's dust-driven wind.

The reported C/O ratios for the Ring Nebula are all less than unity, yet molecules indicative of C-rich gas such as PAHs are present. This is reminiscent of several post-AGB stars that exhibit at the same time O-rich dust and molecules (e.g. crystalline silicates and water ice) and PAHs or fullerenes in their spectra (Waters et al. 1998; Vijh, Witt & Gordon 2005; Gielen et al. 2011; Acke et al. 2013; Malek & Cami 2014). Likewise, Smolders et al. (2010) reported the detection of PAH emission in S-type (equal carbon and oxygen content in atmosphere) AGB stars and attributed the PAH formation in such environments to 'bottom-up' formation from $C_2H_2$, which in turn is formed by non-equilibrium, shock-driven chemistry in the outer regions of the stellar outflow.

**5 CONCLUSIONS**

We present *JWST* NIRSpec IFU and *JWST* MIRI/MRS observations centred on two knots in the Ring Nebula to investigate the PAH emission.

We detect the 3.3 µm PAH feature, and confirm the previous detection of the 11.2 µm PAH feature. There is potentially excess emission between 3.35 and 3.5 µm, which, if born out, indicates the presence of a sizable fraction of PAHs with aliphatic side groups. While the *JWST* observations do not allow for a conclusive detection of the PAH features in the 6–9 µm range, we do detect the 6.2 and 7.7 µm PAH features in *Spitzer*/IRS SL archival observations. The 11.2 µm feature is broad and redshifted, similar to how this feature appears in the Horsehead Nebula.

The *Spitzer*/IRS PAH emission in the 6–9 µm range relative to the 11.2 µm PAH band indicates a population of largely neutral PAHs. The 3.3/11.2 PAH feature ratio indicates small- to medium-sized PAHs, consisting of approximately $35 \pm 6$ C atoms.

We investigate the spatial morphology of the PAH and $H_2$ emission. We find that the 3.3 µm PAH emission peaks in the northern part of the FOV of the north region and is not correlated with the $H_2$ emission probing clumps. Combined with the derived gas density based on the PAH integrated surface brightness ratios, we conclude that the PAH emission originates from the inter-clump region. We further confirm the presence of a ring of enhanced PAH emission centred on the central star reported by Wesson et al. (2024) based on *JWST* Imaging observations. Because the PAHs are located in an outer ring and the region of enhanced $H_2$ emission is inside this outer ring, we propose that the PAHs originate from the AGB wind of the central source.

**ACKNOWLEDGEMENTS**

We are very grateful to Greg Sloan for useful discussions and providing the spectrum of SMP LMC 058. This work is based on observations made with the NASA/ESA/CSA *JWST*. The data were obtained from the Mikulski Archive for Space Telescopes at the Space Telescope Science Institute, which is operated by the Association of Universities for Research in Astronomy, Inc., under NASA contract NAS 5-03127 for *JWST*. These observations are associated with programme ID #1558. This work is based in part on observations made with the *Spitzer Space Telescope*, which was operated by the Jet Propulsion Laboratory, California Institute of Technology under a contract with NASA. EP and JC acknowledge support from the University of Western Ontario, the Institute for Earth and Space Exploration, the Canadian Space Agency (CSA, 22JWGO1-14), and the Natural Sciences and Engineering Research Council of Canada. NLJC and JB-S contributed to this work in the framework of the Agence Nationale de la Recherche's LabCom INCLASS (ANR-19-LCV2-0009), a joint laboratory between ACRI-ST and the Institut d'Astrophysique Spatiale (IAS). MM and RW acknowledge support from the STFC consolidated grant (ST/W000830/1). KJ acknowledges the support from Chalmers University of Technology, Department of Space, Earth and Environment, Onsala Space Observatory, 439 92 Onsala, Sweden. AM acknowledges the support from the State Research Agency (AEI) of the Ministry of Science, Innovation and Universities (MICIU) of the Government of Spain, and the European Regional Development Fund (ERDF), under grants PID2020-115758GB-I00/AEI/10.13039/501100011033 and PID2023-147325NB-I00/AEI/10.13039/501100011033. RS's contribution to the research described here was carried out at the Jet Propulsion Laboratory, California Institute of Technology, under a contract with NASA, and was funded in part by NASA/STScI award JWST-GO-01558.002-A. MJB acknowledges support from European Research Council (ERC) Advanced Grant number SNDUST 694520. This article is based upon work from COST Action CA21126 – Carbon molecular nanostructures in space (NanoSpace), supported by COST (European Cooperation in Science and Technology).

**DATA AVAILABILITY**

*JWST* raw data are available from MAST (programme ID 1558).

The extracted spectra are also archived in MAST: https://archive.stsci.edu/doi/resolve/resolve.html?doi=10.17909/c8k8-hz68.

## APPENDIX A: INFLUENCE OF CONTINUUM DETERMINATION ON PAH STRENGTHS

We aim to assess the influence of the continuum determination on the derived PAH band strengths. We have determined lower and upper bounds for the continuum emission (Figs A1 and A2) and subsequently determined PAH band strengths employing these alternative continua. These PAH band strengths are used to set the uncertainty on the PAH band strengths.

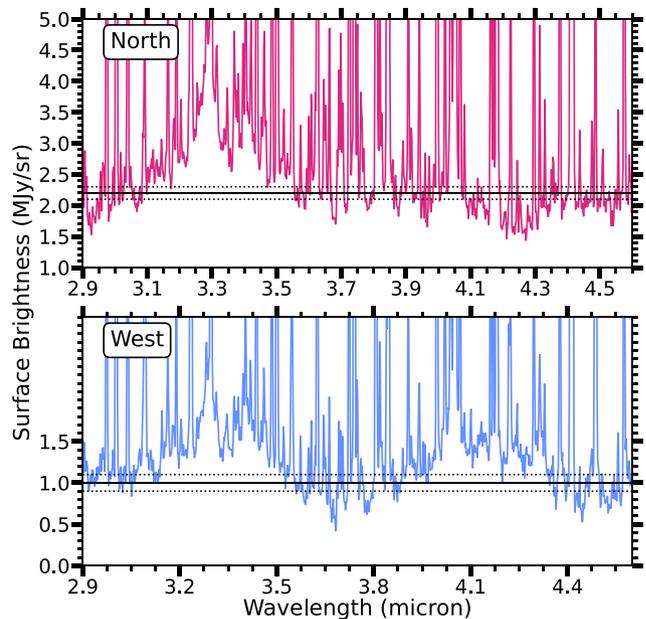

**Figure A1.** Illustration of the continuum assessment for the 3–3.6 μm range. The continuum employed in this paper is presented by a solid line. Upper and lower bounds are shown with dashed lines.

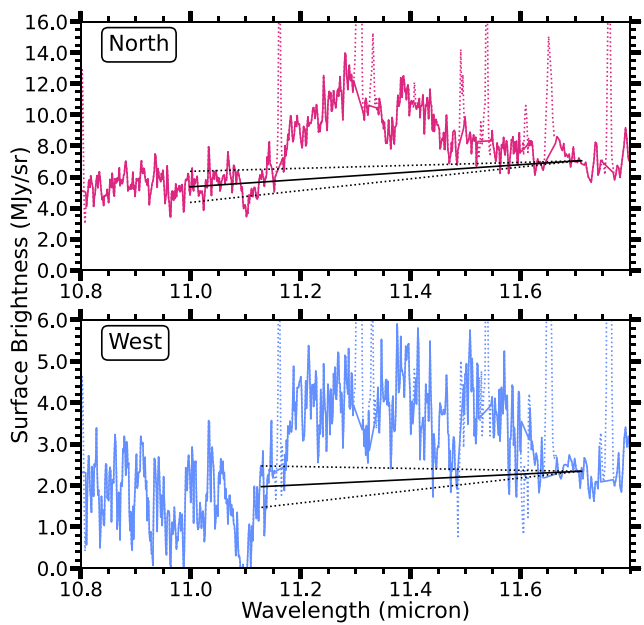

**Figure A2.** Illustration of the continuum assessment for the 11.1–11.6 μm range. The continuum employed in this paper is presented by a solid line. Upper and lower bounds are shown with dashed lines.

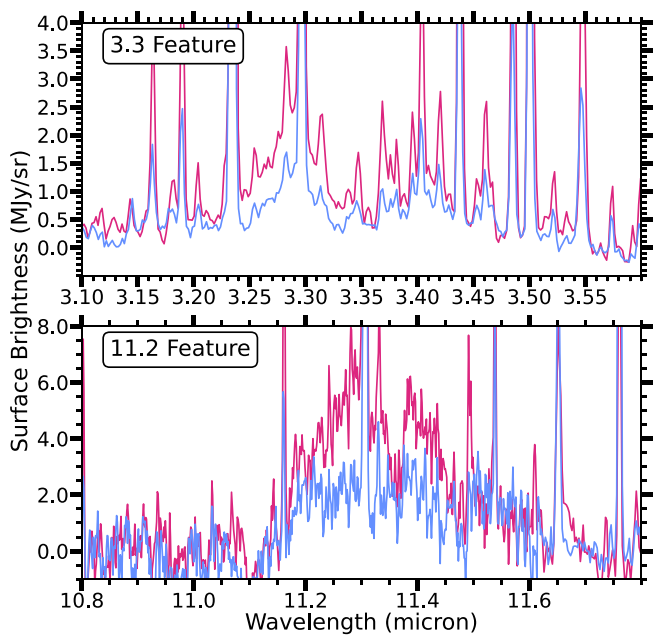

**Figure B1.** A direct comparison of the PAH spectra in the north (pink) and west (blue) regions, after continuum subtraction. The features have similar profiles between the north and west regions. The north region is roughly twice as strong as the west region.

## APPENDIX B: COMPARISON OF THE NORTH AND WEST REGIONS

Fig. B1 shows a direct comparison between the 3.3 μm feature and the 11.2 μm feature of the north and west regions. There are no notable differences in the profiles of these features, the primary difference being their strength.

## APPENDIX C: *SPITZER*/IRS SL OBSERVATIONS AND DATA REDUCTION

We use *Spitzer*/IRS observations identified with the Astronomical Observation Request (AOR) key of 17592832 and 17593088 for, respectively, the on-source and background observations with G. Fazio as PI. The raw data were processed with the S18.18.0 pipeline version by the *Spitzer* Science Center. The resulting basic calibrated data (BCD) products are further processed using cubism (Smith et al. 2007).[9] We used dedicated background observations to correct the background contribution to the data. We applied *cubism*'s automatic bad pixel generation with $\sigma_{TRIM} = 7$ and Minbad-fraction = 0.50 and 0.75 for the global and record bad pixels, respectively.

## APPENDIX D: *JWST* OBSERVATIONS AND DATA REDUCTION FOR THE HORSEHEAD NEBULA, SMP LMC 058, AND NGC 7027

*JWST* observations of the Horsehead Nebula were taken from MAST (Programme ID 1192; PI: K. Misselt) and reduced using the *JWST* reduction pipeline version 1.12.5 and *JWST* CRDS version 11.17.6 and jwst_1197.pmap CRDS context.

*JWST* observations of SMP LMC 058 were taken from MAST (Programme ID 1049; PI: M. Mueller) and reduced using the *JWST* pipeline version 1.15.1 and *JWST* CRDS version 11.17.25 and jwst_1293.pmap CRDS context. A modified pipeline was run by Greg Sloan that treated the 12 channel/grating combinations separately and applied a residual fringe correction to each.

*JWST* observations of NGC 7027 were taken from MAST (Programme ID 1593; PI: D. Law) and reduced using the *JWST* pipeline version 1.15.1 and *JWST* CRDS version 11.18.4 and jwst_1293.pmap CRDS context.

## APPENDIX E: ADDITIONAL MORPHOLOGY PLOTS

In order to increase the confidence of the near-IR morphology results, we re-binned the NIRSpec data to a 2 × 2 grid (Fig. 9). Here, we present the morphology based on the original pixel scale of 0.1 arcsec spaxels (Fig. E1). Despite decreased signal-to-noise ratio level, these morphologies provide further confirmation of the findings in Section 3.3: (i) the $H_2$ emission is structured and traces clumps in the nebula (see also Wesson et al. 2024); and (ii) the PAH emission is clearly displaced from the $H_2$ emission (peaking at larger distances from the central star) and is thus not co-located with the clumps.

We also present signal-to-noise ratio maps for both the original (Fig. E2) and regridded (Fig. E3) and morphology maps. While the signal-to-noise ratio of the original maps is acceptable overall, we regridded the data none the less in order to increase confidence in our results, and because the knot can still be discerned after doing so.

[9] https://irsa.ipac.caltech.edu/data/SPITZER/docs/dataanalysistools/tools/cubism/

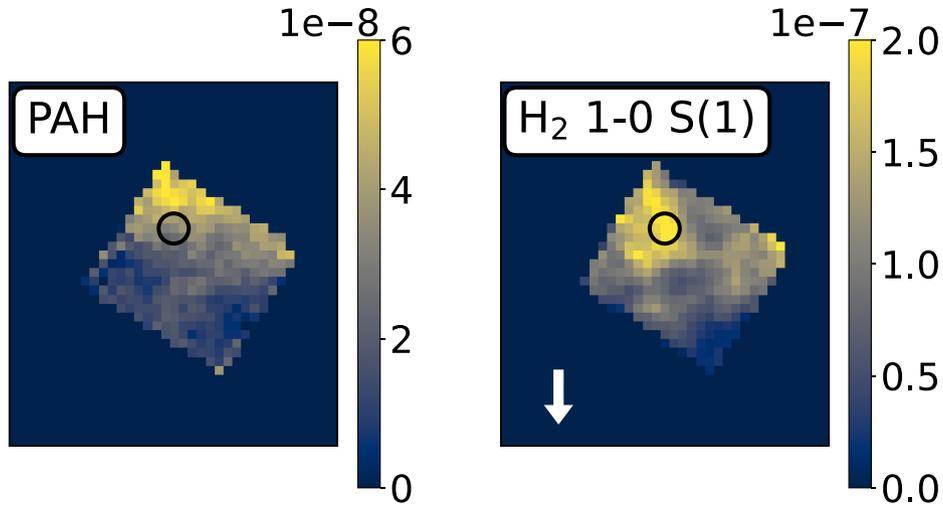

**Figure E1.** The morphology of the 3.3 µm PAH emission (left) and the $H_2$ 1–0 S(1) emission (right) for the north region. The black circle indicates the location of the peak $H_2$ 1–0 S(1) emission. The arrow in the right panel indicates the approximate direction of the central source. This map is in the original pixel scale of 0.1 arcsec spaxels. The integrated surface brightness displayed on the colourbar is in units of $W\,m^{-2}\,sr^{-1}$.

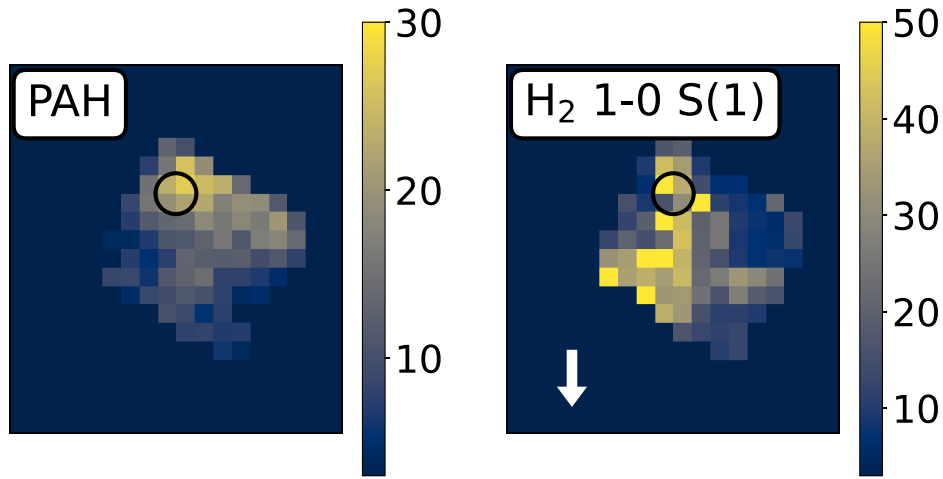

**Figure E2.** The signal-to-noise ratio maps corresponding to Fig. 9 (i.e. when the data are re-binned to a 2 × 2 grid). The black circle indicates the location of the peak $H_2$ 1–0 S(1) emission. The arrow in the right panel indicates the approximate direction of the central source.

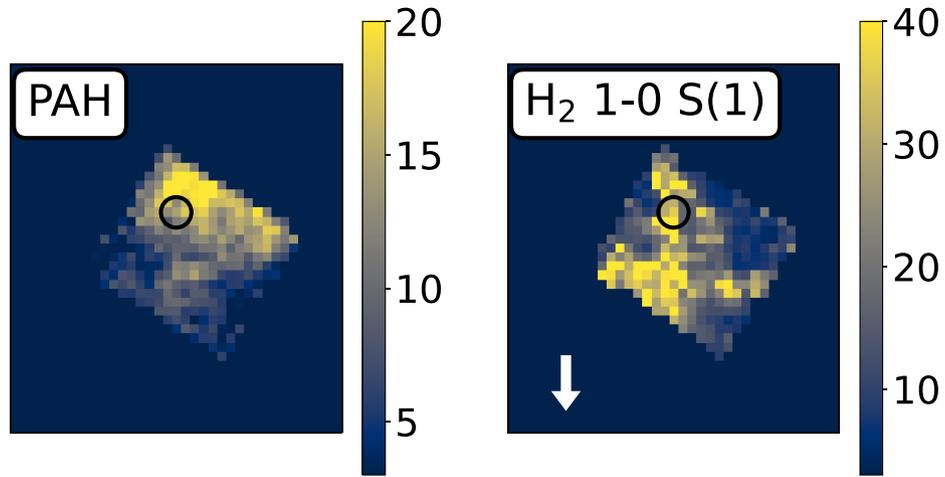

**Figure E3.** The signal-to-noise ratio maps corresponding to Fig. E1 (i.e. on the original pixel scale of 0.1 arcsec spaxels). The black circle indicates the location of the peak $H_2$ 1–0 S(1) emission. The arrow in the right panel indicates the approximate direction of the central source.

This paper has been typeset from a TeX/LaTeX file prepared by the author.